\documentclass[seceq]{ptptex}
\usepackage{graphicx,amsmath}
\newcommand{\pa}{\partial}
\newcommand{\nn}{\nonumber}

\newcommand{\ep}{\epsilon}
\newcommand{\Cbar}{{\bar C}}
\newcommand{\cbar}{{\bar c}}

\title{Ward-Takahashi identity for Yang-Mills theory\\ in the Exact
  Renormalization Group} 
\author{Yuji Igarashi${}^a$, Katsumi
  Itoh${}^a$, Hidenori Sonoda${}^b$}
\inst{
${}^a$ Faculty of Education, Niigata University, Niigata 950-2181, Japan\\
${}^b$ Physics Department, Kobe University, Kobe 657-8501, Japan
}

\abst{We give a functional derivation of the Ward-Takahashi identity
  for Yang-Mills theory in the framework of the exact renormalization
  group.  The identity realizes non-abelian gauge symmetry
  nontrivially despite the presence of a momentum cutoff.  The cutoff
  deforms the gauge transformation by introducing composite operators.
  In our functional method, which is an extension of the method used
  in our previous work on QED, these composite operators are expressed
  in terms of the Wilson action that depends on both a UV cutoff and
  an IR cutoff.}

\begin{document}

\maketitle

\section{Introduction}

Exact renormalization group (ERG)\cite{Wilson}\tocite{Wetterich} has
found applications in a variety of fields and is regarded as a
powerful tool to elucidate non-perturbative physics.\footnote{Recent
  reviews on the ERG are, for example, refs.~\citen{Litim}--\citen{Delamotte}.}
Despite its many successes, there remains an important question in its
application to a field theory with symmetry, in particular, gauge
symmetry.
A momentum cutoff introduced in the ERG approach is often in conflict
with the gauge symmetry present in a field theory.  There are mainly two
different approaches to this problem.  One is to construct a manifestly
gauge invariant regularization scheme.\footnote{Refs. \citen{Morris} and
\citen{Morris2} provide a summary of the approach and perturbative
results.}  The other is to introduce identities for the Green functions
that constrain symmetry breaking terms induced by the momentum
cutoff. In the latter approach, the identity is often written for the
Legendre effective action, $\Gamma$, the generating functional of the
1PI part of the connected cutoff Green functions.  The identity for
$\Gamma$, called the modified Slavnov-Taylor identity
\cite{Ellwanger,Bonini}, has been extensively used to discuss gauge
symmetries in ERG.\footnote{See, for example,
\citen{Pawlowski}, \citen{Gies}, and references therein.} However, symmetry
breaking terms in this identity involve the inverse of the full
propagator, and are consequently somewhat complicated.
For the discussion of symmetry, it becomes more convenient to use the
Ward-Takahashi (WT) identity for the Wilson action $S$, the generating
functional of the cutoff connected Green functions. This is because
the nontrivial Jacobian of the generalized gauge transformation takes
a simple algebraic form; cancellation between the Jacobian and
symmetry breaking terms in $S$ can be seen easily.  The WT identity is
suitable to demonstrate the presence of exact gauge symmetries in ERG.

Recently, one of the present authors (H.S.) derived the WT identity
for QED \cite{sonoda1}.  The identity was subsequently elevated to the
quantum master equation (QME) \cite{igarashi3}, showing the presence
of a nilpotent BRST symmetry, or gauge symmetry, in a cutoff theory.
Furthermore, we have obtained the master action, a solution to the
QME, in terms of the Wilson action satisfying the WT identity.  A more
direct derivation of the master action for QED has also been given in
\citen{HIK}.
The progress described above is a concrete realization of the generic
idea proposed in ref.~\citen{igarashi1} for symmetry realization in ERG.
It is important to stress that the results obtained in
\citen{sonoda1}--\citen{HIK} can be readily generalized to global as well
as local symmetries that are realized linearly in the continuum
limit.\footnote{The general expression of the WT identity for linear
symmetries on the lattice has recently been given in \citen{BBP} as a
generalization of the Ginsparg-Wilson relation\cite{GW} for lattice chiral
symmetry. Related problems are considered in Ref. \citen{igarashi2}.}  Apparently, the presence of QME is not restricted to linear
symmetries such as QED. It is natural to expect that we can extend the
works \citen{sonoda1}--\citen{HIK} to theories with nonlinear (gauge)
symmetry such as Yang-Mills theory.
The WT identity for the Wilson action $S$ in Yang-Mills (YM) theory
was given by Becchi in his pioneering work \cite{Becchi}. This was
actually the first step towards finding exact symmetries in cutoff
theories within the framework of perturbation theory. He introduced
the notion of composite operators to describe generic nonlinear terms
appearing in generalized BRST transformations.  One of the present
authors (H.S.) has recently elaborated on Becchi's program by showing
how to solve QME perturbatively for YM theory
\cite{sonoda2}.\footnote{See also \citen{SU} where QME is solved
  perturbatively for the Wess-Zumino model.  Supersymmetry
  transformation is nonlinear in the absence of auxiliary fields.}

The purpose of this note is to give a functional derivation of the
generic WT identity for nonlinear gauge symmetries, and to apply the
results concretely to the pure SU(2) YM theory. For simplicity, we do
not introduce source terms or anti-fields that generate the BRST
transformation. In contrast to \citen{Becchi} and \citen{sonoda2}, the
BRST transformations we derive are given by composite operators
expressed in terms of functional derivatives of the Wilson action, and
our result is more similar to the WT identity for QED given in
\citen{sonoda1}.

In what follows, the Wilson action is obtained by a functional
integral over fields of momenta between the usual IR cutoff $\Lambda$
and an UV cutoff $\Lambda_0 \gg \Lambda$.  The necessity of the UV
cutoff becomes more apparent when we deal with nonlinear symmetries.
A finite $\Lambda_0$ is necessary to write down the WT identity
explicitly, even though $\Lambda_{0}$ can be eventually sent to
infinity as the theory is renormalizable.  Our approach should be
compared with that given in \citen{Becchi, sonoda2}, where the BRST
invariance of the continuum limit $(\Lambda_{0} \to \infty)$ is used
as an input to obtain the identity at a finite value of an IR cutoff
$\Lambda$.  These two approaches differ in the choice of initial
conditions for the ERG differential equation, and they are physically
equivalent in the limit $\Lambda_0 \to \infty$.

In sect.~2, we consider a generic theory with BRST symmetry, 
and explain our functional derivation of
the WT identity as well as the BRST transformation for the IR theory.
Then, in sect.~3, we apply the results to the pure SU(2) YM theory.

\section{A general functional integral derivation of
      the WT identity}

Consider an action ${\cal S}[\phi]$, a functional of fields $\phi^A$.
The theory is assumed to have some symmetry, that is written as a BRST
symmetry.  The Grassmann parity for $\phi^{A}$ is expressed as
$\ep(\phi^{A})=\ep_{A}$: $\ep_{A} =0~(1)$ if the field $\phi^{A}$ is
Grassmann even (odd).  When we consider a gauge theory, ${\cal
  S}[\phi]$ is the gauge fixed action, and $\phi^A$ represent
collectively gauge and matter fields as well as ghosts, antighosts,
and B-fields.  The index $A$ represents the Lorentz indices of vector
fields, the spinor indices of fermions, and indices distinguishing
different types of generic fields.

In order to regularize the theory, we introduce an IR momentum cutoff
$\Lambda$ and a UV cutoff $\Lambda_{0} > \Lambda$ through a positive
function that behaves as
\begin{eqnarray}
 \quad K\Bigl(\frac{p^{2}}{\Lambda^{2}}\Bigr) ~\simeq~ \left\{
		\begin{array}{ll}
		 1 & (p^2 < \Lambda^2)~, \\
		 0 & (p^2 >  \Lambda^2)~.
		\end{array}
               \right.
\label{cutoff-func}
\end{eqnarray}
In the following, we use two functions $K(p)\equiv K(p^{2}/\Lambda^{2})$ and 
$K_{0}(p)\equiv K(p^{2}/\Lambda_{0}^{2})$.

By introducing sources $J_A$, the generating functional is written as
\begin{eqnarray}
{\cal Z}_{\phi}[J] = \int {\cal D} \phi 
\exp\left(-{\cal S}[\phi ; \Lambda_{0}]+ K_{0}^{-1} J \cdot \phi
       \right ),
\label{part-func1}
\end{eqnarray}
where the action ${\cal S}$ defined at the scale $\Lambda_{0}$ is
written as the sum of the kinetic and interaction terms
\begin{eqnarray}
{\cal S}[\phi; \Lambda_{0}] 
= \frac{1}{2}\phi \cdot K_{0}^{-1} D \cdot \phi + 
{\cal S}_{I}[\phi; \Lambda_{0}]\,.
\label{micro-action}
\end{eqnarray}
In this paper we use the matrix notation in momentum space:
\begin{eqnarray}
J \cdot \phi&=& \int \frac{d^{4}p}{(2\pi)^4}J_{A}(-p)\phi^{A}(p), \nn\\
\phi \cdot  D  \cdot \phi &=& \int \frac{d^{4}p}{(2\pi)^4} 
\phi^{A}(-p) D_{AB}(p) \phi^{B}(p).
\label{cond-not}
\end{eqnarray}

In performing the functional integral (\ref{part-func1}), we decompose
the fields $\phi^A$ with the propagator
$K_{0}(p)\left(D_{AB}(p)\right)^{-1}$ into the sum of the IR fields
$\Phi^A$ with the propagator $K(p)\left(D_{AB}(p)\right)^{-1}$ and the
UV fields $\chi^A$ with $(K_{0}(p)-K(p))\left(D_{AB}(p)\right)^{-1}$.
Note that $\Phi^A$ carry the momenta below $\Lambda$, and $\chi^A$
between $\Lambda_0$ and $\Lambda$.  The integration over the UV fields
$\chi^A$ gives us the generating functional for the IR fields
$\Phi^A$:\cite{igarashi3}
\begin{equation}
{Z}_{\Phi}[J] = \int {\cal D} \Phi 
\exp\left(-{S}[\Phi; \Lambda]+ J \cdot K^{-1}\Phi
       \right )\,,
\end{equation}
where
\begin{equation}
S[\Phi; \Lambda] \equiv \frac{1}{2}\Phi \cdot K^{-1}D \cdot \Phi 
+ {S}_{I}[\Phi; \Lambda]
\label{part-func2}
\end{equation}
is the Wilson action, and its interaction part $S_I$ is defined by
\begin{eqnarray}
\exp \left( -S_{I}[\Phi; \Lambda]\right) \equiv \int{\cal D}\chi
\exp \left[ - \frac{1}{2} \chi \cdot (K_{0}-K)^{-1}D \cdot \chi -
{\cal S}_{I}[\Phi + \chi; \Lambda_{0}] \right]~.
\label{Wilsonian}
\end{eqnarray} 

Two generating functionals (\ref{part-func1}) and
(\ref{part-func2}) are related as
\begin{eqnarray}
{\cal Z}_{\phi}[J]= N_{J} Z_{\Phi}[J],
\label{Z-relation}
\end{eqnarray}
where the normalization factor $N_{J}$ is given by
\begin{eqnarray}
\ln N_{J} &=& -\frac{(-)^{\ep_{A}}}{2} 
J_{A}  K_{0}^{-1}K^{-1}(K_{0}-K) \left(D^{-1}\right)^{AB}  J_{B}~.
\label{normalization}
\end{eqnarray}

We next consider how the symmetry realization is affected by the
presence of cutoffs.  We write the BRST transformation with an
anticommuting constant $\lambda$ as
\begin{eqnarray}
\phi^{A} \rightarrow \phi^{\prime A} =\phi^{A} +  \delta_{\lambda} \phi^{A}~,
\qquad  \delta_{\lambda} \phi^{A} = \delta \phi^{A} \lambda =
\mathcal{R}^{A}[\phi; \Lambda_{0}]~\lambda~.\label{BRST1}
\end{eqnarray}
 
The generating functional (\ref{part-func1}) is invariant under the
change of the integration variable by the BRST transformation
(\ref{BRST1}).  This trivial observation produces the relation
\begin{eqnarray}
\int {\cal D}\phi 
\Bigl(K_0^{-1} J \cdot \delta \phi - \Sigma [\phi; \Lambda_0]\Bigr) 
~\exp\left(-{\cal S}[\phi ; \Lambda_{0}]
+K_{0}^{-1} J \cdot \phi\right) = 0 
\label{equality}
\end{eqnarray}
where $\Sigma [\phi; \Lambda_0]$ is the WT operator given as
\begin{eqnarray}
\Sigma[\phi; \Lambda_{0}] 
 \equiv 
\frac{\pa^{r} {\cal S}}{\pa \phi^{A}} \delta \phi^{A} 
- \frac{\pa^{r} }{\pa \phi^{A}} \delta \phi^{A}~. 
\label{WTop1}
\end{eqnarray} 
$\Sigma [\phi, \Lambda_0]$ is the sum of the change of the original
gauge fixed action ${\cal S}[\phi ; \Lambda_{0}]$
\begin{eqnarray}
\delta_{\lambda} {\cal S} = \frac{\pa^{r} {\cal S}}{\pa \phi^{A}}
\delta_{\lambda} \phi^{A}\,,
\label{delta-cal-S}
\end{eqnarray} 
and that of the functional measure ${\cal D} \phi$
\begin{eqnarray}
\delta_{\lambda} \ln {\cal D} \phi = (-)^{\ep_{A}}
 \frac{\pa^{r} }{\pa \phi^{A}} \delta_{\lambda} \phi^{A}
=\frac{\pa^{r} }{\pa \phi^{A}} \delta \phi^{A}\lambda\,.
\label{measure}
\end{eqnarray} 
The relation (\ref{equality}) may be rewritten as
\begin{eqnarray}
\left\langle\Sigma [\phi ; \Lambda_{0}]\right\rangle_{\phi, ~K_{0}^{-1}J}
&=&
K_0^{-1} J \cdot \left<\delta \phi \right\rangle_{\phi,
~K_{0}^{-1}J} \nonumber\\
&=&
K_0^{-1} J \cdot \left\langle \mathcal{R} [\phi ;
    \Lambda_{0}]\right\rangle_{\phi, ~K_{0}^{-1}J} 
\nn\\
&=&
K_0^{-1} J \cdot \mathcal{R} [K_0 \partial_J^l ; \Lambda_{0}]~{\cal Z}_{\phi}[J]
\label{equality2}
\end{eqnarray}
where the field $\phi^A$ is replaced by the functional derivative with
respect to $J^A$, and $\mathcal{R}^A [\partial_J^l]$ is called the Slavnov
operator.  Note that (\ref{equality2}) is valid whether or not the
theory is invariant under the BRST transformation (\ref{BRST1}).

For an anomaly-free renormalizable theory, we assume that 
the operator (\ref{WTop1}) behaves as
$$
\Sigma[\phi; \Lambda_{0}] = O(1/\Lambda^{2}_{0} )\,,
$$
for a large but finite value of $\Lambda_{0}$.
To be more precise, we assume 
\begin{equation}
\left\langle \Sigma [\phi; \Lambda_0] \right\rangle_{\phi, K_0^{-1} J}
= O (1/\Lambda_0^2)\label{WT-Sigma-phi1}
\end{equation}
or equivalently,
\begin{equation}
K_0^{-1} J \cdot \mathcal{R} [K_0 \partial^l_J; \Lambda_0] \mathcal{Z} [J]
= O (1/\Lambda_0^2)
\end{equation}
for an arbitrary source $J$.\footnote{Later in eq.~(\ref{UV-act}), we
  write the gauge fixed action $S[\phi ; \Lambda_0]$ explicitly.
  Thanks to renormalizability, this action has only a finite number of
  parameters.  Our assumption is that we can tune the parameters so
  that eq.~(\ref{WT-Sigma-phi1}) holds.}  This is the
  statement of the WT identity for the original theory
  defined at the UV scale $\Lambda_0$.

We now wish to transform the above into an equivalent
  condition on the Wilson action $S$ with a finite IR cutoff
  $\Lambda$.  As a preparation let us first summarize important
  properties of composite operators in the ERG framework.  In general,
  given an operator $\mathcal{O} [\phi; \Lambda_0]$ at the UV scale
  $\Lambda_0$, we can define the corresponding IR composite operator
  $\mathcal{O} [\Phi; \Lambda]$ by
\begin{eqnarray}
&&\mathcal{O} [\Phi; \Lambda] \exp \left(- S_I [\Phi; \Lambda]\right)
\equiv \int \mathcal{D} \chi\, \mathcal{O} [\Phi + \chi; \Lambda_0]
\nn\\
&&\qquad \cdot \exp \left[ - \frac{1}{2} \chi \cdot (K_0 - K)^{-1} D
    \cdot \chi - \mathcal{S}_I [\Phi + \chi; \Lambda_0] \right]\,.
\label{defO}
\end{eqnarray}
This operator has two important properties:
\begin{enumerate}
\item The $\Lambda$ dependence is given by the ERG flow equation:
\begin{equation}
\dot{\mathcal{O}} = \frac{\pa^{r}{\cal O}}{\pa \Phi^{A}}
\left({\dot K}D^{-1}\right)^{AB}\frac{\pa^{l} S_{I}}{\pa \Phi^{B}}
-(-)^{\ep_{A}(\ep_{\cal O} +1)}\left({\dot K}D^{-1}\right)^{AB}
\frac{\pa^{l}\pa^{r}{\cal O}}{\pa \Phi^{B}\pa \Phi^{A}}\,,
\label{RGCO}
\end{equation}
where the dot denotes the logarithmic derivative $\Lambda
\frac{\partial}{\partial \Lambda}$.
\item The expectation value in the presence of an arbitrary source satisfies
\begin{equation}
\left\langle \mathcal{O} [\Phi; \Lambda]\right\rangle_{\Phi, K^{-1} J}
= N_J^{-1} \left\langle \mathcal{O} [\phi;
    \Lambda_0] \right\rangle_{\phi, K_0^{-1} J} \,. \label{vevO}
\end{equation}
\end{enumerate}

Now, we apply (\ref{defO}) and (\ref{vevO}).  Using
  (\ref{defO}) we first define the WT operator $\Sigma [\Phi;
  \Lambda]$ for the IR theory by
\begin{eqnarray}
&&\Sigma [\Phi; \Lambda] \exp \left( - S_I [\Phi; \Lambda]\right)
\equiv \int
\mathcal{D} \chi\, \Sigma [\chi+\Phi; \Lambda_0]\nn\\
&&\qquad \times \exp \left[ -
    \frac{1}{2} \chi \cdot (K_0 - K)^{-1} D \cdot \chi - \mathcal{S}_I
    [\Phi + \chi; \Lambda_0] \right]\,.\label{IRWT}
\end{eqnarray}
Then, (\ref{vevO}) implies
\begin{equation}
\left\langle \Sigma[\Phi; \Lambda]\right\rangle_{\Phi, K^{-1} J}
= N_J^{-1} \left\langle \Sigma[\phi; \Lambda_0]\right\rangle_{\phi, K_0^{-1} J}\,.
\label{twoSigmas}
\end{equation}
We also define the IR operator $\delta \Phi^A$ corresponding to
$\delta \phi^A$ by
\begin{eqnarray}
    &&\delta \Phi^A [\Phi; \Lambda] \exp \left(- S_I [\Phi; \Lambda] \right)
    \equiv K K_0^{-1} \int \mathcal{D} \chi\, \delta \phi^A [\Phi + \chi;
    \Lambda_0]\nn\\
    &&\qquad \times \exp \left[ -
        \frac{1}{2} \chi \cdot (K_0-K)^{-1} D \cdot \chi -
        \mathcal{S}_I [\Phi+\chi; \Lambda_0] \right] \label{deltaPhi}
\end{eqnarray}
(\ref{vevO}) implies
\begin{equation}
\left\langle \delta \Phi^A \right\rangle_{\Phi, K^{-1} J}
= K K_0^{-1} N_J^{-1} \left\langle \delta \phi^A \right\rangle_{\phi, K_0^{-1} J}
\end{equation}
Replacing the field by a functional derivative with respect to a
source, we obtain
\begin{equation}
R^A [K \partial^l_J; \Lambda] Z_\Phi [J] = K K_0^{-1} N_J^{-1} \mathcal{R}^A
[K_0 \partial^l_J; \Lambda_0] \mathcal{Z}_\phi [J]\label{twoRs}
\end{equation}
where we denote
\begin{equation}
\delta \Phi^A [\Phi;\Lambda] = R^A [\Phi; \Lambda]\,.
\end{equation}
Now, using (\ref{equality2}), we obtain
\begin{equation}
N_J^{-1} \left\langle \Sigma [\phi; \Lambda_0]\right\rangle_{\phi, K_0^{-1} J}
= N_{J}^{-1}K_{0}^{-1}J 
\cdot \mathcal{R} [K_{0}\pa^{l}_J; \Lambda_{0}] ~\mathcal{Z}_{\phi}[J]\,.
\end{equation}
Hence, using (\ref{twoSigmas}) and (\ref{twoRs}), we obtain
\begin{equation}
\left\langle \Sigma [\Phi; \Lambda]\right\rangle_{\Phi, K^{-1} J}
= K^{-1} J \cdot R[K \partial^l_J; \Lambda] Z_\Phi [J]\,.
\end{equation}
This implies
\begin{equation}
\Sigma [\Phi; \Lambda] = \frac{\partial^r S [\Phi; \Lambda]}{\partial
  \Phi^A} \delta \Phi^A - \frac{\partial^r}{\partial \Phi^A} \delta
\Phi^A \,.
\label{IRWTbyS}
\end{equation}
This relation is analogous to the original WT operator (\ref{WTop1}),
but note here that $S$ is the Wilson action at an IR scale $\Lambda$,
and that $\delta \Phi^A$ are composite operators expressed in terms of
$S$.  For $\Lambda = \Lambda_0$, $\Sigma [\Phi; \Lambda]$ becomes the
original $\Sigma [\phi; \Lambda_0]$, as is obvious from the
definition.

Consequently, the WT identity for the Wilson action $S$ is given
either as
\begin{equation}
\Sigma [\Phi; \Lambda] = O (1/\Lambda_0^2)
\end{equation}
or
\begin{equation}
K^{-1} J \cdot R[K \partial^l_J; \Lambda] Z_\Phi [J] = O (1/\Lambda_0^2)\,.
\end{equation}

Before concluding this section, we consider a
  particular class of BRST transformation
\begin{eqnarray}
\delta \phi^{A} =
\mathcal{R}^{A}[\phi; \Lambda_{0}] = K_0
\Bigl(\mathcal{R}^{(1)A}_{~~~B} (\Lambda_{0}) ~\phi^B  
+ \frac{1}{2} \mathcal{R}^{(2)A}_{~~~BC} (\Lambda_{0})~\phi^B \phi^C \Bigr)~.
\label{form of BRST}
\end{eqnarray}
which are at most quadratic in fields.  The BRST transformation for YM
theories belongs to this class.

In rewriting the expectation value 
\[
\langle \delta \phi^A
\rangle_{\phi, K_0^{-1}J} = \mathcal{R}^A [K_0 \partial^l_J; \Lambda_0]~{\cal
  Z}_{\phi}[J]
\] 
for the IR theory, we need to compute two things.  First,
\begin{eqnarray}
    K_{0} \frac{\pa^{l}}{\pa J_{A}} {\cal Z}_{\phi}[J]
    &=& K_{0}\frac{\pa^{l}}{\pa J_{A}}
    N_{J}Z_{\Phi}[J]\nn\\
    &=& N_{J} \left[(-)^{\ep_{A} +1}
        \left(\frac{(K_{0}-K)}{K}D^{-1}\right)^{AB}J_{B} +
        K_{0}\frac{\pa^{l}}{\pa J_{A}}\right]   Z_{\Phi}[J]
    \nn\\
    &=& N_{J} \left\langle K_{0}K^{-1}\Phi^{A} - (K_{0}-K)\left(D^{-1}\right)^{AB}
        \frac{\pa^{l} S }{\pa \Phi^{B}}\right\rangle_{\Phi,K^{-1}J}\nn\\
    &=& N_{J}\left\langle \Phi^{A} - (K_{0}-K)\left(D^{-1}\right)^{AB}
        \frac{\pa^{l}S_{I} }{\pa \Phi^{B}}\right\rangle_{\Phi,K^{-1}J}\nn\\
    &\equiv& N_{J}\left\langle\left[\Phi^{A}\right]_{\rm
          com}\right\rangle_{\Phi,K^{-1}J},  
\label{CO1}
\end{eqnarray}
where we define
\begin{eqnarray}
\left[\Phi^{A}\right]_{\rm com} &\equiv& K_{0}K^{-1}\Phi^{A} 
- (K_{0}-K)\left(D^{-1}\right)^{AB}
\frac{\pa^{l} S }{\pa \Phi^{B}}\nn\\
&=& \Phi^{A} - (K_{0}-K)\left(D^{-1}\right)^{AB}
\frac{\pa^{l} S_{I}}{\pa \Phi^{B}}. 
\label{def-CO}
\end{eqnarray}
Second,
\begin{eqnarray}
K_{0}^{2} \frac{\pa^{l}}{\pa J_{A}}\frac{\pa^{l}}{\pa J_{B}}
 {\cal Z}_{\phi}[J]
&=& K_{0}^{2} \frac{\pa^{l}}{\pa J_{A}}
\frac{\pa^{l}}{\pa J_{B}}
N_{J}Z_{\Phi}[J]\nn\\
&\equiv& N_{J} \left\langle
\left[\Phi^{A}~\Phi^{B}\right]_{\rm com}\right\rangle_{\Phi,K^{-1}J}~,
\label{CO2}
\end{eqnarray}
where we define
\begin{eqnarray}
\Bigl[\Phi^{A}~\Phi^{B}\Bigr]_{\rm com} 
&\equiv& \left[\Phi^{A}\right]_{\rm com}~
\left[\Phi^{B}\right]_{\rm com}\nn\\
&{}& \hspace{-6mm} - (K_{0}-K)\left(D^{-1}\right)^{AC}
(K_{0}-K)\left(D^{-1}\right)^{BD}
\frac{\pa^{l} \pa^{l}S_{I}}{\pa \Phi^{C} \pa \Phi^{D}}\,.
\label{CO3}
\end{eqnarray}
It is important to note that nontrivial contributions arise from
derivatives $\partial_J$ acting on the normalization factor $N_{J}$.

Hence, using (\ref{twoRs}), we obtain
\begin{equation}
\left\langle \delta \Phi^A\right\rangle_{\Phi, K^{-1} J} = K \left\langle
\mathcal{R}^{(1) A}_{~~~B} (\Lambda_0) \left[\Phi^A\right]_{\mathrm{com}}
+ \frac{1}{2} \mathcal{R}^{(2) A}_{~~~BC} (\Lambda_0) \left[\Phi^A
    \Phi^B\right]_{\mathrm{com}}\right\rangle_{\Phi, K^{-1} J}\,.
\end{equation}
Since this is valid for arbitrary $J$, we obtain the operator equality
\begin{eqnarray}
\delta \Phi^A = K \Bigl(
\mathcal{R}^{(1)A}_{~~~B} (\Lambda_0) \left[\Phi^B\right]_{\rm com} 
+ \frac{1}{2} \mathcal{R}^{(2)A}_{~~~BC} (\Lambda_0) \left[\Phi^B
    \Phi^C\right]_{\rm com} \Bigr)~. 
\label{IR BRST tf}
\end{eqnarray} 
It is important to stress the necessity of the cutoff function $K_{0}$
to make (\ref{IR BRST tf}) UV finite.\footnote{The potential UV
divergence is hard to see in the matrix notation.  It is hidden in the
loop momentum integral contained in $\mathcal{R}^{(2)A}_{~~~BC} [\Phi^B
\Phi^C]_{\mathrm{com}}$.  In the next section, where we apply the above
results to a pure YM theory, we will elaborate on this point. See the
remark right after (\ref{BRST5}).}

\section{Pure Yang-Mills theory: WT identity and BRST transformation}

Let us find the explicit form of $\Sigma[\Phi ; \Lambda]$ for the SU(2) 
pure Yang-Mills theory.  We use the following notations:
\[
\begin{array}{r@{\quad}l}
\textrm{UV~ fields}&\phi^{A} \equiv \{a^{a}_{\mu}, ~b^{a}, ~c^{a}, ~
\cbar^{a}\}\nn\\
\textrm{Source~ terms}&J_{A} \equiv
 \{J^{a}_{\mu},~J^{a}_{B},~J^{a}_{c},~J^{a}_{\cbar} \}
\end{array}
\]
As a UV action we take\footnote{For the SU(2) group adjoint 
indices we
  use the notation $A \cdot B = A^{a} B^{a}$ and $(A \times B)^{a} =
  \epsilon^{abc} A^{b} B^{c}$.}
\begin{eqnarray}
{\cal S}[\phi ; \Lambda_{0}]= 
\frac{1}{2}\phi \cdot K_{0}^{-1} D \cdot \phi + {\cal S}_{I}[\phi]
\label{UV-act}
\end{eqnarray}
where
\begin{eqnarray}
\frac{1}{2}\phi \cdot K_{0}^{-1} D \cdot \phi &=& \int_p K_{0}^{-1}(p) 
\Bigl[\frac{1}{2}
 a_{\mu}(-p) \cdot 
(p^2 \delta_{\mu\nu}-  p_{\mu}p_{\nu}) a_{\nu}(p) 
\nn\\
&&  + \cbar(-p)ip^{2}
\cdot c(p) -b(-p) \cdot 
\bigl(ip_{\mu}a_{\mu}(p) + \frac{\xi}{2}b(p)\bigr)\Bigr]\nn
\end{eqnarray}
and
\begin{eqnarray}
&& \hspace{-1cm} 
{\cal S}_{I}[\phi ; \Lambda_{0}] = 
 \int_{p} \biggl[\frac{a_{2}}{2}\Lambda_{0}^{2} a_{\mu}(-p)a_{\mu}(p)
+ \frac{z_{1}}{2} p^{2} a_{\mu}(p)a_{\mu}(-p)
+ \frac{z_{2}}{2} p_{\mu} p_{\nu} a_{\mu}(-p) a_{\nu}(p) \biggr]\nn\\
&& \qquad
+z_{3} \int_{p,q} p_{\nu} a_{\mu}(-p) \cdot 
[a_{\mu}(q) \times a_{\nu}(p-q)]
\nn\\
&& \qquad
 +\frac{z_{4}}{8} \int_{p_{1},\cdot\cdot,p_{4}} \delta\Bigl(\sum p_{i}\Bigr)
a_{\mu}(p_{1})\cdot a_{\mu}(p_{2})a_{\nu}(p_{3})\cdot a_{\nu}(p_{4})\nn\\
&& \qquad
  +\frac{z_{5}}{8} \int_{p_{1},\cdot\cdot,p_{4}} \delta\Bigl(\sum p_{i}\Bigr)
 a_{\mu}(p_{1})\cdot a_{\nu}(p_{2})a_{\mu}(p_{3})\cdot a_{\nu}(p_{4})\nn\\
&& \qquad
- \int_{p} \biggl[  p_{\mu} \cbar(-p)
\cdot \Bigl( (-i)z_{6}p_{\mu}c(p) +z_{7}
\int_{q} a_{\mu}(p-q)\times c(q)
 \Bigr)\biggr]\,.
\nn
\end{eqnarray} 

The BRST transformation is given by
\begin{eqnarray}
\delta a_{\mu}(p) &=& K_0 (p) \left[ (1 + z_{6} K_0 (p) ) (-i) p_\mu ~c(p) + 
z_{7} K_0 (p) \int_{q}a_{\mu}(p-q)\times c(q)\right] ,\nn\\ 
\delta \cbar(p) &=& i K_{0}(p)~b(p) ,\nn\\ 
\delta c(p) &=& K_{0}(p)\frac{z_{8}}{2}~\int_{q}c(q)\times c(p-q).
\label{BRST3}
\end{eqnarray}
where we have chosen $\delta a_\mu (p)$ so that
\begin{equation}
p_\mu \delta a_\mu (p) = - K_0 (p)^2 \frac{\partial^l}{\partial \bar{c}
  (-p)} \mathcal{S} [\phi; \Lambda_0]
\end{equation}
is the operator for the ghost equation of motion.  Note $z_8$ is an
independent parameter that does not appear in the action.  We tune the
eight $z$ coefficients to satisfy the WT identity.\footnote{The
logarithmic dependence of $z$'s is determined by renormalizability.  It
is the part independent of $\Lambda_0$ that must be tuned.}

Let us compute 
\begin{eqnarray}
    \left(K_{0}^{-1}J \cdot \mathcal{R} [K_{0}{\pa^{l}_J}, \Lambda_0] \right)
    {\cal Z}_{\phi}[J] = 
    N_{J}^{-1}\left(K_{0}^{-1}J \cdot \mathcal{R} [K_{0}{\pa^{l}_J},
        \Lambda_0] \right) N_{J}Z_{\Phi}[J]
\label{WT-YM1}
\end{eqnarray}
where the Slavnov operator is given by
\begin{eqnarray}
K_{0}^{-1}J \cdot \mathcal{R} [K_{0}{\pa^{l}_J}, \Lambda_{0}] 
&=& \int_{p} J_{\mu}(-p)\cdot 
\biggl\{(1+z_{6}K_{0}(p)) (-ip_{\mu}) K_0 (p) \frac{\pa^{l}}{\pa J_{c}(-p)}\nn\\
&{}&
\hspace{-1.5cm}
 + z_{7} K_0 (p) \int_{k}K_{0}(p-k)K_{0}(k)\frac{\pa^{l}}{\pa
   J_{\mu}(-p+k)} \times  
\frac{\pa^{l}}{\pa J_{c}(-k)}\biggr\}\nn\\
&{}&
\hspace{-1.5cm}
+\int_{p}\biggl[ J_{\cbar}(-p)\cdot i~K_{0}(p)
\frac{\pa^{l}}{\pa J_{B}(-p)}
\label{slavnov1}\\
&{}&
\hspace{-1.5cm}
+ \frac{z_{8}}{2}J_{c}(-p)\cdot \int_{k}K_{0}(p-k)K_{0}(k)
\frac{\pa^{l}}{\pa J_{c}(-p+k)}\times 
\frac{\pa^{l}}{\pa J_{c}(-k)}
\biggr]~.
\nn
\end{eqnarray}
For the pure YM case, the source dependent normalization factor $N_J$
takes the form
\begin{eqnarray}
\ln N_{J} &=& -\frac{(-)^{\ep_{A}}}{2} 
J_{A} \Bigl(\frac{K_{0}-K}{K_{0}K}\Bigr) \left(D^{-1}\right)^{AB} J_{B}\nn\\
&=& \int_p \Bigl(\frac{K_{0}-K}{K_{0}K}\Bigr)(p) 
\Bigl\{
J_{c}(-p) \frac{-i}{p^2} J_{\bar c}(p) 
- J_{B}(-p) \frac{-ip_{\mu}}{p^2} J_{\mu}(p)\nn\\
&{}&~~~~~~- \frac{1}{2}J_{\mu}(-p) \frac{1}{p^2} 
\Bigl(\delta_{\mu\nu}-(1-\xi)\frac{p_{\mu}p_{\nu}}{p^2}\Bigr) 
J_{\nu}(p)
\Bigr\}~.
\label{nf}
\end{eqnarray} 

It is easy to see that the derivatives with respect to $J_A$ in
(\ref{slavnov1}) give rise to composite operators.  The following
composite operators appear in the WT identity:
\begin{eqnarray}
&{}& \left[A_{\mu}(p)\right]_{\rm com} \equiv A_{\mu}(p)-
\frac{K_{0}(p)-K(p)}{p^{2}} 
\left(\delta_{\mu\nu}- (1-\xi)\frac{p_{\mu}p_{\nu}}{p^{2}}\right)
\frac{\pa S_{I}}{\pa A_{\mu}(-p)}~,\nn\\ 
&{}& \left[B(p)\right]_{\rm com} = B(p) + i~ p_{\mu}
\frac{K_{0}(p)-K(p)}{p^{2}}\frac{\pa S_{I}}{\pa A_{\mu}(-p)}~, 
\nn\\
&{}&\left[C(p)\right]_{\rm com} \equiv C(p)+i \frac{K_{0}(p)-K(p)}{p^{2}}
\frac{\pa^l S_{I}}{\pa \Cbar(-p)}~,\nn\\  
&{}&\left[A_{\mu}(q) \times C(p-q)
\right]_{\rm com} \equiv \left[A_{\mu}(q)\right]_{\rm com}\times
\left[C(p-q)\right]_{\rm com}~ \nn\\
&{}&\qquad +i \frac{K_{0}(q)-K(q)}{q^{2}}
\left(\delta_{{\mu}{\nu}}- (1-\xi)\frac{q_{\mu}q_{\nu}}{q^{2}}\right)\nn\\
&{}&\qquad\qquad\qquad \cdot \frac{K_{0}(p-q)-K(p-q)}{(p-q)^{2}}
~\Bigl(\frac{\pa^{l}}{\pa A_{\nu}(-q)} \times
\frac{\pa^{l} S_{I}}{\pa \Cbar(-p+q)}\Bigr)~,
\nn\\
&{}&\left[C(q)\times C(p-q)
\right]_{\rm com} \equiv \left[C(q)\right]_{\rm com}\times
\left[C(p-q)\right]_{\rm com}~ \nn\\
&{}& \qquad +
\frac{K_{0}(q)-K(q)}{q^{2}}\frac{K_{0}(p-q)-K(p-q)}{(p-q)^{2}}
\frac{\pa^{l}}{\pa \Cbar (-q)}\times 
\frac{\pa^{l} S_{I}}{\pa \Cbar(-p+q)}~.
\label{CO3b}
\end{eqnarray} 

After some calculations, we find that the WT identity is given in the
form of eq.~(\ref{IRWTbyS}).  To be concrete, we obtain the WT
identity
\begin{eqnarray}
&{}&\Sigma[\Phi,~\Lambda] \equiv \int_{p} \Bigl(\frac{\pa S}{\pa
  A_{\mu}(p)}\delta A_{\mu}(p) 
 + \frac{\pa^{r} S}{\pa \Cbar(p)}\delta \Cbar(p) + \frac{\pa^{r} S}{\pa
 C(p)}\delta C(p)\nn\\
&{}& ~~~~~~~~ - \frac{\pa}{\pa A_{\mu}(p)} \delta A_{\mu}(p) 
+  \frac{\pa^l}{\pa C(p)} \delta C(p)
\Bigr)= O (1/\Lambda_0^2)
\label{WT-YM5}
\end{eqnarray}
with the BRST transformation given as
\begin{eqnarray}
\delta A_{\mu}(p) &=& K(p)\biggl( -i(1+z_{6} K_0 (p))
p_{\mu}[C(p)]_{\rm com} \nn\\ && \quad 
+ z_{7} K_0 (p) \int_{q}\Bigl[A_{\mu}(q) \times C(p-q)
\Bigr]_{\rm com}\biggr)~, \nn\\
\delta \Cbar(p)&=& i K(p) [B(p)]_{\rm com}~, \nn\\
\delta C(p)&=& \frac{z_{8}}{2}
K(p)\int_{q}\Bigl[C(q)\times C(p-q)\Bigr]_{\rm com}~.
\label{BRST5}
\end{eqnarray}
Here, the integrals over $q$ for $\delta A_\mu$ and $\delta C$ are
  finite, thanks to the $K_{0}$ with the UV cutoff $\Lambda_{0}$.  Note
  that the logarithmic dependence of $z_6, z_7, z_8$ on $\Lambda_0$ is
  chosen to cancel that of these integrals over $q$.  Hence, we can
  define further the composite operators
\begin{eqnarray}
\left[ (A_\mu \times C) (p)\right]_{\mathrm{com}} &\equiv& - i z_6
p_\mu \left[C(p)\right]_{\mathrm{com}} + z_7 \int_q \left[ A_\mu (q)
    \times C(p-q) \right]_{\mathrm{com}}\\
\left[ \left(C \times C\right) (p)\right]_{\mathrm{com}} &\equiv& z_8
\int_q \left[ C(q) \times C(p-q) \right]_{\mathrm{com}}
\end{eqnarray}
which have a limit as $\Lambda_0 \to \infty$.  Then, in this limit we
obtain $\Sigma [\Phi, \Lambda] = 0$ with
\begin{equation}
\left\lbrace
\begin{array}{c@{~=~}l}
\delta A_\mu (p) & K(p) \left( - i [C(p)]_{\mathrm{com}} + \left[
        (A_\mu \times C) (p)\right]_{\mathrm{com}} \right)\\
\delta \bar{C} (p) &  K(p) i \left[B(p)\right]_{\mathrm{com}}\\
\delta C (p) & K(p) \frac{1}{2} \left[ \left(C \times C\right)
    (p)\right]_{\mathrm{com}} 
\end{array}\right.
\end{equation}

\section{Discussion}

Extending our previous work \cite{igarashi3}, we have derived the WT
identity for generic nonlinear gauge symmetries such as those in YM
theory, and have read off the corresponding BRST transformation. Since
the BRST symmetry in YM theory is nonlinear even classically, the
deformation of the symmetry due to the presence of a momentum cutoff
becomes highly nontrivial. In our functional method, the deformation
of the BRST symmetry is described by some contributions generated by
the Slavnov operator acting on the normalization factor which is
quadratic in sources. The Slavnov operator is characterized by the
first and second functional derivatives of sources, and generates
particular combinations of factors with functional derivatives of the
Wilson action. They satisfy the ERG flow equations for the composite
operators. In our functional method, however, we need to introduce a
UV cutoff $\Lambda_{0}$ in addition to an IR cutoff $\Lambda$ to make
the WT identity well-defined. This is the price we pay for insisting
on expressing the nonlinear BRST transformation explicitly in terms of
the Wilson action.  Alternatively, we can construct the composite
operators by using their flow equations without introducing
$\Lambda_{0}$, as discussed in \citen{Becchi} and \citen{sonoda2}. These
two approaches will be equivalent at least within the framework of
perturbation theory.

It should be remarked that even in the QED case, the BRST transformation
for the Wilson action is not nilpotent.  This observation motivated us
to elevate the WT identity to QME in the Batalin-Vilkovisky formalism\cite{BV,BV2} in
our earlier paper \cite{igarashi3}.  Including the antifield
contributions, the nilpotency is recovered, and the BRST invariance of
the system becomes easy to see.  Naturally, we have a similar situation
here for YM theory: the BRST transformation in eq. (\ref{BRST5}) is not
nilpotent.  So our next immediate task is to reformulate the theory in
the BV formalism.

The general ERG formalism guarantees the existence of QME for YM theory.
In fact in the approach by Becchi \cite{Becchi} in which the
renormalized theory is constructed directly without introducing an UV
theory at scale $\Lambda_0$, a quantum master action satisfying QME has
been constructed \cite{sonoda2}.  But Becchi's approach applies only to
perturbation theory.  In a forthcoming paper,\cite{igarashi4} we plan to
construct a quantum master action with both UV and IR cutoffs.  The
advantage of this construction is its applicability beyond perturbation
theory.

\section*{Acknowledgements}

The work by K.~I. was supported in part by the Incentive Grant 2007 from
Department of Education, Niigata University. Y.~I. thanks the Institute of 
Theoretical Physics in Heidelberg for hospitality. He is also grateful 
to J.~M.~Pawlowski for enlightening discussions.


\end{document}